\def\nk{n_{\rm b}}

\def\rfr#1{Eq. (\ref{#1})}

\def\virg#1{``#1"}

\def\eqi{\begin{equation}}
\def\eqf{\end{equation}}
\def\eqia{\begin{eqnarray}}
\def\eqfa{\end{eqnarray}}
\def\Om{\mathit{\Omega}}
\def\rp#1#2{{#1\over#2}}
\def\lb#1{\label{#1}}

\def\ton#1{\left(#1\right)}

\def\grf#1{\left\{#1\right\}}

\documentclass{ws-ijmpd}
\usepackage[super,compress]{cite}
\usepackage[combine,document]{ucs}
\usepackage{url,wasysym}
\usepackage[polutonikogreek,english]{babel}
\usepackage[utf8x]{inputenx}
\usepackage[LGR,T1,T5]{fontenc}
\usepackage[breaklinks]{hyperref}
\hypersetup{colorlinks,urlcolor=blue,citecolor=green,linkcolor=red,filecolor=red}
\usepackage{breakurl}
\usepackage{txfonts}
\usepackage{paralist}
\newcommand{\grk}[1]{\selectlanguage{polutonikogreek}
#1\selectlanguage{english}}
\begin{document}

\markboth{L. Iorio}
{Gravitational Anomalies in the Solar System?}

%
\catchline{}{}{}{}{}
%

\title{GRAVITATIONAL ANOMALIES IN THE SOLAR SYSTEM?
}

\author{LORENZO IORIO
}


\address{CNR-Istituto di metodologie inorganiche e dei plasmi (I.M.I.P), Via Amendola, 122/D\\
Bari, 70126,
Italy\\
lorenzo.iorio@libero.it}

\maketitle

\begin{history}
\received{Day Month Year}
\revised{Day Month Year}
\end{history}

\begin{abstract}
Mindful of the anomalous perihelion precession of Mercury discovered by U. Le Verrier in the second half of the nineteenth century and its successful explanation by A. Einstein with his General Theory of Relativity in the early years of the twentieth century, discrepancies among observed effects in our Solar system and their theoretical predictions on the basis of the currently accepted laws of gravitation applied to known matter-energy distributions  have the potential of paving the way for remarkable advances in fundamental physics. This is particularly important now more than ever, given that most of the universe seems to be made of unknown substances dubbed Dark Matter and Dark Energy. Should this not be directly the case, Solar system's anomalies could anyhow lead to advancements in either cumulative science, as shown to us by the discovery of Neptune in the first half of the nineteenth century, and technology itself. Moreover, investigations in one of such directions can serendipitously enrich the other one as well. The current status of some alleged gravitational anomalies in the Solar system is critically reviewed. They are:
\begin{inparaenum}[\itshape a\upshape)]
\item Possible anomalous advances of planetary perihelia;
\item Unexplained orbital residuals of a recently discovered moon of Uranus (Mab);
\item The lingering unexplained secular increase of the eccentricity of the orbit of the Moon;
\item The so-called Faint Young Sun Paradox;
\item The secular decrease of the mass parameter of the Sun;
\item The Flyby Anomaly;
\item The Pioneer Anomaly; and
\item The anomalous secular increase of the astronomical unit
\end{inparaenum}
\end{abstract}


\ccode{PACS numbers: 95.30.Sf; 04.80.-y; 04.80.Cc; 04.50.Kd; 95.10.Km; 95.55.Pe}


\section{Introduction}\lb{introduzione}
Armed with beautiful and self-consistent theories of ever-increasing sophistication which have previously passed severe empirical scrutinies, scientists are continuously engaged in a close dialogue with Nature to establish how far the validity of their ambitious theoretical constructs does extend. In doing that, they attempt to  mathematically predict the evolutionary course of as many different systems is possible, and put their predictions to the test by comparing them with dedicated measurements and/or observations as accurate as possible. Occasionally, something may go `wrong': anomalies may show up. In science, the word\footnote{From \grk{>anwmal'ia, as, <h}, made of the privative prefix \grk{>a-} and \grk{<omal'os, 'h, 'on} (`average', `regular').} `anomaly' is used to designate some sort of discrepancy with respect to an expected path occurring in a given phenomenon. In astronomical contexts, it was used since ancient times\footnote{Geminus Astronomicus (\textit{Gem}.1.20; cf. \textit{Ptol}.\textit{Alm.}3.3), deals with \grk{>anwmal'ia t~hs kin'hsews} (`irregularity of motion'). Plutarch, in his \textit{Lives} (\textit{Plut.\AE m}.17), mentions \grk{>anwmal'iai >ekleiptika'i} (`irregularities of the Moon's orbit').} to indicate irregularities in  motions of celestial objects. The aforementioned disagreements have always to be intended in a statistically significant sense in light of the state-of-the-art of the observational and/or measurement techniques used; in other words, the size of an alleged anomalous effect must always be greater than the associated uncertainty-which has always to be released-by a stipulated amount. If an anomaly occurs, great care and attention is required since it could be a clue that we might be facing something important. First of all, it must be carefully ascertained if the anomaly is really out there; it may be either a mere artifact of the data reduction procedure, or the consequence of malfunctioning of the measuring devices and systematic errors in the observations. As an example coming from the recent history of astronomy, near the end of the eighties of the past century the rumor spread about alleged `unexplained residuals' in the orbit of Uranus.\cite{1988CeMec..43...55S, 1988CeMec..43..409S} Actually, their size-a fraction of an arcsecond-was just comparable with that of many known sources of systematic errors in the observations themselves; nonetheless, they were used to predict the position of a so far unseen distant planet in the remote peripheries of our Solar
system\cite{1988AJ.....96.1476H,1989Icar...80..334G,1988A&A...203..170G,1989JBIS...42..327P,1992A&A...255..401B}: the time-honored Planet X scenario\cite{PX09} which has often resurfaced so far for disparate, more or less sound theoretical and/or observational reasons. It was later pointed out\cite{1993AJ....105.2000S} that an inaccurate value of the mass of Neptune was used in the dynamical models used to calculate the path of Uranus. Also possible errors in adjusting the orbit of Uranus were pointed out in Ref.~\refcite{1993AJ....105.2000S}. If, after repeated checks and re-analyses of either the previously collected data or of new ones the anomaly still lingers on, then new, exciting scenarios may open up. Let us, now, move to the main character of the aforementioned historical example to illustrate how this can occur.

The gravitational interaction\cite{1996lsg..book.....N} weaves the fabric of the natural world.
\begin{inparaenum}[(i)]
\item It shapes the stages on which unknown forms of life may grow up in alien worlds by necessarily conforming to the structural limits imposed by it;
\item Governs the sidereal course of celestial objects over the eons, from the tiniest humble chunks\footnote{At this level, also non-gravitational
    forces\cite{nongrav} come into play in a non-negligible way.} of matter which may eventually concur to form planets to the largest majestic superclusters of galaxies;
\item Decides the ultimate fate of glowing stars, from the shiniest mighty supergiant to the dimmest starlet;
\item Dictates how time rolls by and space unfolds;
\item Married with quantum mechanics in a form which has still to be fully grasped, it is likely responsible of what happens in the inapproachable depths of the black holes' pits, and it was at work at the birth of the universe.
\end{inparaenum}
Among the known four fundamental interactions of Nature, gravitation is the weakest one, but its effects, experienced by everything is endowed with matter/energy, are cumulative and far-reaching. Thus, since the early epochs of human civilization studying the heavens' courses represented a privileged way to catch its secrets thanks to long observational records which, in the last decades, have been continuously enriched by a wealth of observations of ever-increasing exquisite
accuracy returned by man-made spacecrafts sailing the interplanetary regions of our Solar
system.\cite{1986IAUS..114..329A,Moyer2003,2003Natur.425..374B,2004SSRv..115....1K,2007IJMPD..16.2117I,2009AcAau..65..666I,2014AcAau..94..699I}
To date, the best understanding  of gravitation is provided us by the General Theory of Relativity\cite{1915SPAW.......778E,1916AnP...354..769E} (GTR) of A. Einstein which, on completely different theoretical grounds, extends the previous Newton's theory of Universal Gravitation\cite{Newton1687} to which it reduces in the limit of weak fields and slow motions. For a comprehensive, modern account of GTR, see, e.g., Ref.~\refcite{2010grav.book.....P}.

In the middle of the nineteenth century, two gravitational anomalies as large as dozens of arcseconds unambiguously emerged from long records of optical observations of Uranus\cite{LeVer1846} and Mercury\cite{LeVer1859} accurate at about the arcsecond level. They both arose from an application of the laws of the gravitational interaction universally accepted by then to the major bodies of the Solar system which were known at that time. Nonetheless, they met with somewhat opposite lot. The irregularities in the motion of Uranus, interpreted in terms of the Newton's gravitational law, led to the discovery of the still undetected planet Neptune.\cite{Galle1846} For a historical account, see, e.g., Ref.~\refcite{1979dine.book.....G}. Instead, the anomalous perihelion precession of Mercury\footnote{The value reported by U. Le Verrier was\cite{LeVer1859} 38.3 arcseconds per century (" cy$^{-1}$), later corrected to $41.2\pm 2.1$ " cy$^{-1}$ by S. Newcomb.\cite{Newcomb1895} Modern determinations based on radar ranging led to\cite{1972PhRvL..28.1594S} $43.2\pm 0.9$ " cy$^{-1}$. The latest determination, based on optical data, yields\cite{Bucarest010} $42.8$ " cy$^{-1}$.} could  be finally explained only by Einstein\cite{1915SPAW...47..831E} with its GTR\footnote{His\cite{1915SPAW...47..831E} predicted value amounted to 43 " cy$^{-1}$, later corrected  to $43.03$ " cy$^{-1}$ by G. Clemence.\cite{1947RvMP...19..361C} Using the best accepted values for the astronomical constants and for the orbital elements of Mercury, the GTR expected value is nowadays\cite{1986Natur.320...39N,2003Ap&SS.284.1159P} 42.98 " cy$^{-1}$.}, i.e. by modifying the gravitational laws accepted till then, despite the attempts to repeat the trick of Neptune by keeping the Newtonian laws and postulating the hypothetical planet Vulcan. For historical accounts of the vicissitudes of Mercury's perihelion rate, see, e.g., Refs.~\citen{1982mpfl.book.....R,1997ispv.book.....B}. In both cases, breakthroughs  occurred in our knowledge, although not on the same level since the second one was directly related to a paradigm shift in fundamental physics. After all, whether a further planet (made of standard stuff) existed or not was accidental within the Newtonian framework which would not have been affected if it had not existed at all. Moreover, Neptune is not even made by some form of exotic, non-baryonic matter, despite some speculations about such a possibility\cite{2009PhLB..671..203A}.

The dichotomy exhibited by the Neptune/Mercury case arose again several years later under the appearance of further anomalies in the motions of quite different celestial objects, and it is currently at the forefront of the astrophysical research bridging fields as disparate as particle physics, cosmology and astrophysics. In 1933, F. Zwicky\cite{1933AcHPh...6..110Z} showed that the kinematics of the galaxies forming the Coma Cluster (Abell 1656) could not eventually be explained in terms of the Newtonian laws of gravitation applied to the electromagnetically detectable matter of that galactic clump. Zwicky propounded to solve that puzzle by invoking large amounts of still unseen ordinary matter under the rule of the Newtonian gravitational law. Interestingly, in 1937 Zwicky  \cite{1937PhRv...51..290Z} suggested that such invisible matter distributions  could be mapped by observing galaxies acting as gravitational lenses, a possibility that that came to pass only in the nineties of the twentieth century.\cite{1998ApJ...501..539T} Some decades after the Zwicky's seminal work, in the eighties of the twentieth century, an analogous issue showed up in some spiral galaxies\cite{1981AJ.....86.1791B,1983Sci...220.1339R} whose peripheral stars revolve faster than allowed by the luminous matter detected.\cite{1996MNRAS.281...27P} Contrary to the Neptune case, further studies on primordial nucleosynthesis\cite{2009NJPh...11j5028J,2013PhRvD..87l3530C}, gravitational microlensing\cite{2000ApJ...542..281A,2007A&A...469..387T}, anisotropies\cite{2009ApJS..180..225H} in the Cosmic Microwave Background (CMB), Baryon Acoustic Oscillation (BAO) clustering  in sky surveys\cite{2007MNRAS.381.1053P}, and structure formation\cite{2005Natur.435..629S,2006ApJ...647..201C}  elucidated that galactic and extragalactic obscure mass  cannot be made, to a large extent,  of particles of ordinary matter. This is what is commonly meant today by the denomination  `Dark Matter'. Astoundingly, such an exotic form of substance, which does not interact electromagnetically with baryonic stuff, would amount to\cite{2014A&A...571A..16P} $26.5\%$ of the entire material content of the universe, against a paltry\cite{2014A&A...571A..16P} $4.8\%$ for ordinary matter.
A different line of research, pursued by some scientists, encompasses the possibility that new gravitational laws, both at the Newtonian and relativistic level, are actually required to accommodate the Dark Matter puzzle without invoking unknown forms of matter of which direct detections are still missing. The most famous of such alternatives is, perhaps, the MOND (MOdified Newtonian Dynamics) paradigm proposed by M. Milgrom in the early eighties of the twentieth century.\cite{1983ApJ...270..365M,1983ApJ...270..371M,1983ApJ...270..384M} For recent reviews, see, e.g., Refs.~\citen{2012LRR....15...10F,2014SchpJ...931410M}.  Basically, its main tenets\cite{2014SchpJ...931410M} consist of the introduction of an universal acceleration scale $A_0$, ubiquitous  in galactic systems and the universe at large, below which gravitational laws change suitably their form, and space-time scale-invariance of this low-acceleration limit. MOND, which is a collection of theories satisfying such basic tenets, must face all the independent phenomena in which Dark Matter has given good results so far. While several relativistic formulations of MOND appeared in the last years accounted well for the observed gravitational lensing\cite{2014SchpJ...931410M}, they do not yet provide a satisfactory description of cosmology and structure formation.\cite{2014SchpJ...931410M} Interestingly, a direct connection between the search for (allegedly baryonic) dark matter in our Solar system in the form of one or more trans-Plutonian planets and MOND recently arose.\cite{2010OAJ.....3....1I,2013CeMDA.116..357I,2014MNRAS.444L..78I} Indeed, a certain version\cite{2009MNRAS.399..474M} of the so-called External Field Effect (EFE) characterizing within MOND the dynamics of a gravitationally bound system immersed in an external gravitational field can induce anomalous orbital effects\cite{2011MNRAS.412.2530B,2014PhRvD..89j2002H} which mimics the action of a still undiscovered Planet X located in a specific position in the sky.\cite{2010OAJ.....3....1I,2013CeMDA.116..357I,2014MNRAS.444L..78I}  Latest developments in MOND\cite{2012MNRAS.426..673M} predict peculiar signatures even in absence of EFE which could be put to the test in the deep Newtonian regime characterizing our Solar system by looking for anomalies in orbital motions of its major bodies.\cite{2012MNRAS.426..673M,2013CQGra..30p5018I} Another alternative gravitational paradigm aiming to explain\cite{2006ApJ...636..721B,2006MNRAS.367..527B,2007MNRAS.382...29B,Mof013} the Dark Matter phenomenology without resorting to it is the\cite{2006JCAP...03..004M} Scalar-tensor-vector gravity (STVG), known also as MOdified Gravity (MOG), developed by J.~W. Moffat.
Recently, a nonlocal generalization of the Einsteinian theory of gravitation has been proposed by B. Mashhoon and coworkers\cite{2010PhRvD..81f5020B,2014PhRvD..90l4031M,Mash014}.

The late nineties of the twentieth century witnessed the emergence of a new anomaly at an even larger scale. The analyses of distant Type 1a Supernov{\ae} by two independent teams of astronomers\cite{1998AJ....116.1009R,1998ApJ...507...46S, 1999ApJ...517..565P} led to the conclusion that an acceleration in the expansion of the universe should have begun to occur as from about 5 Gyr ago, contrary to what expected on the basis of the application of the general relativistic field equations to the matter-energy content of the universe commonly accepted until then. Later, the existence of such an anomaly of cosmological nature was  supported by several independent observations over the years.\cite{2012CRPhy..13..521A}From a phenomenological point of view, the accelerated expansion of the universe would be driven by a hypothetical form of energy, dubbed `Dark Energy',  which permeates all of space\cite{2003RvMP...75..559P}. Perhaps even more surprisingly than for Dark Matter, such an entity, whose physical nature is unknown, would constitute\cite{2014A&A...571A..16P} $68.25\%$ of the entire matter-energy content of the universe. Several physical mechanisms have been devised so far for Dark Energy, ranging from a small positive cosmological constant-an energy density filling space homogeneously and unchanging over time\cite{2001LRR.....4....1C}-to various quintessence scenarios\cite{quintess} in which Dark Energy is, in general, allowed to vary in space and time.
Interestingly, latest years have seen an increasing interest in proposing laboratory experiments aimed to test various aspects of Dark Energy\cite{2006JPhCS..31..123B,2006PhLB..641....6M,2009AdAst2009E...4D,2011JCAP...09..020B,2014PhRvL.112o1105J}. Moreover, the  Dvali-Gabadadze-Porrati (DGP) braneworld model\cite{2000PhLB..485..208D}, arisen at the dawn of our century in the framework of multidimensional modified models of gravity to explain the cosmological acceleration and nowadays facing severe theoretical and observational challenges\cite{2006PhRvD..73d4016G,2008PhRvD..78j3509F,2010LRR....13....3D}, predicted, among other things, an anomalous extra-precession of the pericenter\cite{2003PhRvD..67f4002L} of a test particle orbiting a central massive body which is independent of the size of its orbit and whose magnitude is comparable to the present-day level of accuracy in determining planetary orbital motions in our Solar system\cite{2011CeMDA.111..363F,2013AstL...39..141P,2013MNRAS.432.3431P}. In the framework of the so-called dark degeneracy\cite{2014JCAP...06..010C}, attempts to provide a unified picture of Dark Matter and Dark Energy have been formulated, e.g., in terms of the Chaplygin gas\cite{2001PhLB..511..265K,2002PhRvD..66d3507B,2004PhRvD..70h3519B,2012EPJC...72.1883X}. S.~Capozziello and coworkers recently looked at the so-called Extended Theories of Gravity\cite{2011PhR...509..167C,2014PhRvD..90d4052C} as a new paradigm aiming to encompass, in a self-consistent scheme, problems like inflation, Dark Energy, Dark Matter, large scale structure and, first of all, to give at least an effective description of Quantum Gravity.

Have the courses of either anthropogenic or natural bodies in our Solar system currently some surprises in store for us which could be understood as  genuine gravitational anomalies needing explanations? If so, what could be their significance and extent for our understanding of gravitational physics? Is there a \textit{fil rouge} interweaving them? Are there, or there were in the recent past, some false alarms needing deeper scrutiny?  If, on the one hand, GTR is a `rigid' theory in the sense that it does not contain free, adjustable parameters, on the other hand, extra gravitational degrees of freedom
may have an environmental dependence\cite{2004PhRvL..93q1104K,2012PhR...513....1C}. Thus, although it would be arguably naive to expect huge departures from GTR in our Solar system, it can be nonetheless considered as a valid testing bench. Indeed, an increasing number of space-based missions endowed with new measurement devices will provide us with a wealth of more and more accurate data in the near future. Moreover, the systematic errors, either observational or dynamical, affecting our artificial and natural probes and potentially biasing our interpretations are generally known and/or can be modeled relatively well.

A general \textit{caveat} concerning the search for possible explanations to alleged anomalies is in order. They are often sought in the form of accelerations due to any particular physical mechanism  which have just to be of the same order of magnitude of the ones responsible of the putative anomalies of interest. In fact, simply restricting oneself to a mere back-of-the-envelope evaluation would be insufficient to draw meaningful conclusions. Instead,  calculations as detailed as possible should be performed to check if the proposed novel accelerations do induce the specific anomalous effects which have been actually detected.

Finally, as a disambiguation remark, we mention that the term `gravitational anomaly' is used also in theoretical physics, but with  a different meaning with respect to the one intended  here. Indeed, it usually designates breakdown of general covariance in certain quantum field theories due to pathologies in one-loop level Feynman diagrams\cite{1984NuPhB.234..269A,Witten85}.

The plan of the paper is as follows.
In Section~\ref{perieli}, the possible existence of new anomalous perihelion precessions in the Solar system is reviewed.
A putative orbital anomaly affecting one of the recently discovered natural satellites of Uranus is discussed in Section~\ref{MAB}.
Section~\ref{luna} deals with the lingering unexplained secular increase of the eccentricity of the Moon's orbit, despite recent efforts to improve the geophysical models of the intricate tidal phenomena taking place in the interior of our planet and its natural satellite. The Faint Young Sun Paradox, according to which the Sun would have been too faint in the Archean to warm the Earth enough to keep liquid water on its surface despite compelling evidence that, instead, there were oceans at that time, is treated in Section~\ref{fysp} as a potential gravitational anomaly. Section~\ref{massa} is devoted to the recently determined decrease of the gravitational parameter of the Sun from Solar system's planetary motions. The Flyby Anomaly, detected in the asymptotic line-of-sight velocities of certain interplanetary spacecraft during some but not all of their Earth flybys to receive their gravity assists, is the subject of Section~\ref{flyby}. Some of the recent developments about the time-honored issue of the anomalous acceleration experienced by the Pioneer 10 and 11 probes approximately after they passed the $\approx 20$ au threshold, known as Pioneer Anomaly, are discussed in Section~\ref{Pio}. Latest developments about the anomalous secular increase of the astronomical unit are illustrated in Section~\ref{UA}. In Section~\ref{fine}, our conclusions are offered.
\section{Anomalous Perihelion Precessions}\lb{perieli}
As reviewed in Section \ref{introduzione}, the successful explanation of the long-standing puzzle posed by the anomalous Hermean\footnote{From \grk{<Erm~hs} (`Hermes'), identified with the Roman deity Mercury.} perihelion precession was a milestone in the history of the empirical corroborations of GTR.\cite{Sha80} Could  history repeat itself, with new orbital anomalies paving the way to the possible discovery of modifications of the currently accepted general relativistic laws of the gravitational interaction?

Nowadays, GTR is fully included\cite{Esta71, Moyer00} in either the dynamical and measurement multiparameteric models routinely fit to huge data sets in the building process of the modern ephemerides (DE, INPOP, EPM) at the first post-Newtonian (1PN) level by treating the major bodies of the Solar system as point particles, i.e. by neglecting their rotation. It implies that the 1PN gravitomagnetic fields of the Sun and the other bodies of the Solar system, causing the Lense-Thirring orbital precessions\cite{LT18}, are neglected; see Section~\ref{leti}. Usually, the relativistic equations are expressed in terms of the parameters $\beta$ and $\gamma$ of the parameterized
post-Newtonian (PPN) formalism\cite{1968PhRv..169.1017N, 1971ApJ...163..611W, 1972ApJ...177..757W, 1993tegp.book.....W}, which, in some global solutions of the teams producing the ephemerides, are treated as solve-for parameters estimated along with hundreds of other ones. The teams responsible of the INPOP (Institut de m\'{e}canique c\'{e}leste et de calcul des \'{e}ph\'{e}m\'{e}rides) and EPM (Institute of Applied Astronomy of the Russian Academy of Sciences) ephemerides for several years are independently producing global solutions in which also corrections $\Delta\dot\varpi$ to the standard planetary  precessions of the longitudes of perihelia\footnote{The longitude of perihelion $\varpi\doteq\Om+\omega$ is a `dogleg' angle since it is the sum of two angles, i.e. the longitude of the ascending node $\Om$ and the argument of pericenter $\omega$, lying in different planes\cite{1999ssd..book.....M}.} $\varpi$ are determined, with different methodological approaches, by keeping $\beta$ and $\gamma$ fixed to their relativistic values, i.e. $\beta=\gamma=1$. The most recent values are listed in Table \ref{ta1}.
 \begin{table}
\tbl{Corrections $\Delta\dot\varpi$ to the standard secular perihelion precessions of some planets of the Solar system, in milliarcseconds per century (mas cy$^{-1}$), phenomenologically determined by contrasting with different methodologies the dynamical and measurement models of the recent ephemerides INPOP10a\cite{2011CeMDA.111..363F} and EPM2011\cite{2013AstL...39..141P, 2013MNRAS.432.3431P} to different data records spanning almost one century. Planetary dynamics was generally modeled, to different levels of completeness, by including several known Newtonian effects (N-body perturbations including those from the major asteroids and some Trans-Neptunian Objects, Sun's oblateness, ring of minor asteroids) and GTR to the first post-Newtonian (1PN) level with the exception of the Solar gravitomagnetic field causing the Lense-Thirring precessions.\cite{LT18} The supplementary precessions $\Delta\dot\varpi$ globally account for any statistical and systematic inaccuracies in our knowledge of the Solar system's dynamics, the propagation of the electromagnetic waves, the functioning of the measuring devices, in the observations and in the data reduction process itself. Any anomalous perihelion precessions, caused by whatsoever modification of the current laws of gravity one can devise,  must conform with such ranges of admissible values. Note that the extra-precessions of Venus and Jupiter, obtained with the EPM2011 ephemerides, are statistically compatible with non-zero effects\cite{2013AstL...39..141P, 2013MNRAS.432.3431P}, although at a modest level of significance:$1.6\sigma$ (Venus), $2\sigma$ (Jupiter).}
{\begin{tabular}{@{}ccccccc@{}} \toprule
Ephemeris & Mercury & Venus & Earth & Mars & Jupiter & Saturn \\
 & (mas cy$^{-1}$) & (mas cy$^{-1}$) & (mas cy$^{-1}$) & (mas cy$^{-1}$) & (mas cy$^{-1}$) & (mas cy$^{-1}$) \\ \colrule
EPM2011\cite{2013AstL...39..141P, 2013MNRAS.432.3431P} & $-2.0\pm 3.0$ & $2.6\pm 1.6$ & $0.19\pm0.19$ & $-0.020\pm 0.037$ & $\hphantom{0}58.7\pm 28.3$ & $-0.32\pm 0.47$ \\
INPOP10a\cite{2011CeMDA.111..363F}\hphantom{0} & $\hphantom{0}0.4\pm 0.6$ & $0.2\pm 1.5$ & $-0.20\pm 0.90\hphantom{0}$ & $-0.040\pm 0.150$ & $-41.0\pm 42.0$ & $\hphantom{0}0.15\pm 0.65$ \\
\botrule
\end{tabular} \label{ta1}}
\end{table}
In this case, the adjective `standard' is referred not only to the Newtonian laws, as it was at the time of Le Verrier, but also to the relativistic ones in the aforementioned sense. As such, any `anomalous' perihelion precession that might be discovered, would not have anything to do with the historical anomalous perihelion precession of Mercury of $42.98$ " cy$^{-1}$ since it is nowadays fully included in  the state-of-the-art  models of all of the modern ephemerides. Instead, if real, it would be due to some unmodelled dynamical effects which, in principle, could potentially signal a breakthrough in the currently accepted laws of gravitation. In any case, the figures of Table \ref{ta1} represent the current ranges of admissible values for any anomalous perihelion precession, and were determined in a phenomenological way, independently of any particular alternative model of gravity.

A few years ago, it seemed that something strange was going on with the orbit of Saturn. Indeed, in a private exchange with the present author occurred in October 2008, the astronomer E.~V. Pitjeva mentioned a non-zero retrograde precession of the Kronian\footnote{From \grk{Kr'onos} (`Cronus/Cronos/Kronos'), identified with the Roman deity Saturn.} perihelion as large as\cite{2009AJ....137.3615I} \eqi\Delta\dot\varpi_{\saturn} = -6\pm 2~\textrm{mas~cy}^{-1}\lb{pitpre}\eqf obtained by processing some Cassini data with the EPM2008 ephemerides\cite{Pit09}. Such a result, mentioned also in Ref.~\refcite{Fienga010} in which Ref.~\refcite{Pit09} is cited, was independently confirmed by a team led by the astronomer A. Fienga\cite{Fienga010} who analysed Cassini data with the INPOP08 ephemerides obtaining a marginally significant ($1.25\sigma$) extra-precession \eqi\Delta\dot\varpi_{\saturn} = -10\pm 8~\textrm{mas~cy}^{-1}.\lb{fiepre}\eqf Although at different levels of statistical significance, both \rfr{pitpre} and \rfr{fiepre} are mutually compatible. The result of \rfr{pitpre} was discussed in Ref.~\refcite{2009AJ....137.3615I}, where some possible explanations in terms of some conventional and unconventional gravitational effects were sought. The orbit of Saturn exhibited seemingly anomalous features also in the DE ephemerides produced by the Jet Propulsion Laboratory, although no supplementary perihelion precessions have ever been explicitly determined so far by its team of astronomers currently led by W.~M. Folkner. Indeed, the residuals of right ascension $\alpha$, declination $\delta$ and range $\rho$ of Saturn,  produced with the DE421 ephemerides and Cassini tracking data from 2005 to 2007, showed  unambiguous anomalous patterns\cite{DE421}. Later investigations with both the INPOP\cite{2011CeMDA.111..363F} and the EPM\cite{Pit010,2013AstL...39..141P,2013MNRAS.432.3431P} ephemerides disproved the existence of the anomalous perihelion precession of Saturn, whose latest determinations turned out to be statistically compatible with zero (See Table \ref{ta1}). Also further data analyses with the DE430 ephemerides by
JPL\cite{IPN014,2014PhRvD..89j2002H} did not confirm the anomalous signature in the Earth-Saturn range by offering also an explanation of the previously reported anomalies. According to Ref.~\refcite{2014PhRvD..89j2002H}, they were due to an incorrect reduction of the Cassini data in which spacecraft orbits relative to Saturn were fit to both the Doppler and range measurements, causing the sinusoidal signature in the Earth-Saturn range residuals of, e.g., Fig. 7 in Ref.~\refcite{2014PhRvD..89j2002H}.  Actually, the resulting spacecraft trajectories in separate orbits segments were not independent because of the use of the ranging data in the trajectory fits. This led to the approach adopted in Refs.~\citen{IPN014,2014PhRvD..89j2002H} of fitting the spacecraft trajectories by using only the Doppler range-rate data, allowing the range measurements to be used to determine the orbit of Saturn.
The outcome of the analysis in Ref.~\refcite{2014PhRvD..89j2002H}, primarily aimed to constrain a peculiar form of\cite{2009MNRAS.399..474M} EFE within the MOND framework, can be transposed in terms of a correction to the Kronian perihelion precession\cite{2014PhRvD..89j2002H}
\eqi\Delta\dot\varpi_{\saturn} = 0.43\pm 0.43~\textrm{mas~cy}^{-1}\lb{heespre}\eqf
which is statistically compatible with zero, in agreement with the values in Table \ref{ta1}.

An examination of Table \ref{ta1} reveals that, according to the EPM2011 ephemerides, the Cytherean\footnote{From \grk{K'ujhra} (`Cytherea'), identified with the Roman deity Venus.} and Jovian perihelia  might actually undergo anomalous precessions\cite{2013AstL...39..141P, 2013MNRAS.432.3431P}, although the level of statistical significance is somewhat modest. Further data analyses are required to confirm or disproof such putative anomalies. In Ref.~\refcite{2014Galax...2..466A}, a preliminary explanation is offered in terms of a non-standard form of transversal gravitomagnetism generated by the Sun. If they will turn out to be genuine physical effects requiring explanation, the availability of more than one non-zero perihelion extra-precession has the potential of being used to put effectively to the test any modified model $\mathcal{M}$ of gravity which predicts additional rates of change $\dot\varpi_{{\mathcal{M}}}$ for such an orbital element. Indeed, the ratio
\eqi {\mathcal{C}}_{{\mathcal{M}}}^{(\textrm{A,B})} \doteq\rp{\dot\varpi^{(\textrm{A})}_{{\mathcal{M}}}}{\dot\varpi^{(\textrm{B})}_{{\mathcal{M}}}}\eqf of such theoretically predicted precessions for a given pair of planets A, B is independent of the multiplicative parameter  which usually characterizes the model $\mathcal{M}$ of interest, depending only on the orbital geometries of the planets themselves. Thus, ${\mathcal{C}}_{{\mathcal{M}}}^{(\textrm{A,B})}$ could be compared with the observationally-based ratio
\eqi {\mathcal{O}}^{(\textrm{A,B})} \doteq\rp{\Delta\dot\varpi^{(\textrm{A})}}{\Delta\dot\varpi^{(\textrm{B})}}\eqf
of the non-zero corrections $\Delta\dot\varpi$ to the standard perihelion precessions for the same planets A, B. Such an approach was proposed for the first time in Ref.~\refcite{2007AdHEP200790731I}, although incorrectly implemented since corrections $\Delta\dot\varpi$ statistically compatible with zero were used.
In the case of Venus and Jupiter, from Table \ref{ta1} a Root-Sum-Square (RSS) error propagation yields
\begin{align}
{\mathcal{O}}^{(\textrm{\venus,\jupiter})} \lb{primo} & = 0.044\pm 0.034, \\ \nonumber \\
{\mathcal{O}}^{(\textrm{\jupiter,\venus})} \lb{secondo} & = 22.57\pm 17.65.
\end{align}
Such values can be used with a number of theoretically predicted extra-precessions of the perihelion. As an example, a cosmological constant $\Lambda$ affects a test particle which revolves about a central body of mass $M$ making its ellipse of semimajor axis $a$ and eccentricity $e$ secularly precess at a  rate\cite{2003CQGra..20.2727K}
\eqi\dot\varpi_{\Lambda}=\rp{1}{2}\ton{\rp{\Lambda c^2}{\nk}}\sqrt{1-e^2},\lb{ccos} \eqf
where $c$ is the speed of light in vacuum, $\nk\doteq\sqrt{GM a^{-3}}$ is the planetary Keplerian mean motion, and $G$ is the Newtonian constant of gravitation. As such, the theoretical ratio for \rfr{ccos} is \eqi\mathcal{C}_{\Lambda}^{(\textrm{A,B})}=\sqrt{\rp{a^3_{\textrm{A}}\ton{1-e^2_{\textrm{A}}}}{a^3_{\textrm{B}}\ton{1-e^2_{\textrm{B}}}}},\lb{theo}\eqf
which is independent of $\Lambda$.
In the case of Venus and Jupiter, \rfr{theo} yields
\begin{align}
{\mathcal{C}}_{\Lambda}^{(\textrm{\venus,\jupiter})} \lb{terzo} & = 0.051, \\ \nonumber \\
{\mathcal{C}}_{\Lambda}^{(\textrm{\jupiter,\venus})} \lb{quarto} & = 19.27.
\end{align}
It can be noted that \rfr{terzo}-\rfr{quarto} are compatible with \rfr{primo}-\rfr{secondo}; it would not be so if the uncertainties of the supplementary precessions of Venus and Jupiter in Table \ref{ta1} were smaller by one order of magnitude, with potentially relevant consequences for our currently accepted views on the cosmological acceleration and Dark Energy. This example should adequately illustrate  the importance of having trustable and independently cross-checked observations and data reduction procedures when alleged anomalies come into play.

At present, the orbit determination of the two outermost planets of the Solar system and the dwarf planet Pluto are mostly based on optical observations of relatively modest accuracy, if compared to usual radiometric ones. Thus, no particularly accurate corrections\cite{PitPio,Pit010} $\Delta\dot\varpi$ have been determined so far, also because the available data records cover barely a full period of Uranus and only parts of the orbits of Neptune and Pluto: indeed, the current accuracies in their $\Delta\dot\varpi$ are just at the\cite{PitPio,Pit010} " cy$^{-1}$  level (cfr. with the figures in Table \ref{ta1} for the other planets).  They are displayed in Table \ref{ta2}.
 \begin{table}
\tbl{Corrections $\Delta\dot\varpi$ to the standard secular node precessions of Uranus, Neptune and Pluto, in " cy$^{-1}$, phenomenologically determined by contrasting in a least-square sense the dynamical and measurement models of the ephemerides EPM2008\cite{Pit010} to data records spanning almost one century. See Table \ref{ta1} for the inner planets.}
{\begin{tabular}{@{}cccc@{}} \toprule
Ephemeris & Uranus & Neptune & Pluto  \\
 & (" cy$^{-1}$) & (" cy$^{-1}$) & (" cy$^{-1}$) \\ \colrule
EPM2008\cite{Pit010}\hphantom{0} &  $-3.89\pm 3.90$ & $-4.44\pm 5.40 $ & $2.84\pm 4.51$\\
\botrule
\end{tabular} \label{ta2}}
\end{table}
Such limitations could be overcome with some proposed missions to Uranus (Uranus Pathfinder\cite{2012ExA....33..753A}) and Neptune (Outer Solar System-OSS\cite{2012ExA....34..203C}), and the New Horizons spacecraft\cite{2008SSRv..140....3S}, currently en route to the Plutonian system and equipped with advanced transmitting apparatuses to perform accurate radio-science experiments.\cite{2008SSRv..140..217T}  About the gaseous giants, in the course of the selection of the scientific themes for the second and third L-class missions of the Cosmic Vision 2015-2025 program of the European Space Agency (ESA), their exploration  was encouragingly defined \virg{a timely milestone, fully appropriate for an L class mission}.  In Ref.~\refcite{2014P&SS..104...93T}, a single L-class mission dedicated to both planets was discussed. Long records of accurate radiometric data for so remote objects  would be of paramount importance for accurate tests of gravitational theories in the distant outskirts of the Solar system as well (See Section~\ref{Pio}).

As a final consideration, if and when the teams of astronomers will begin to systematically determine corrections to the standard secular rates of changes  of all the other orbital elements, it will be possible to perform more independent gravitational tests by taking into account also alternative theories  predicting non-spherically symmetric corrections to the standard Newtonian potential. At present, only the INPOP team released corrections\cite{2011CeMDA.111..363F} $\Delta\dot\Om$ to the node precessions obtained with the INPOP10a ephemerides which are, in general, less accurate than those for the perihelia; they are shown in Table \ref{ta3}.
 \begin{table}
\tbl{Corrections $\Delta\dot\Om$ to the standard secular node precessions of some planets of the Solar system, in milliarcseconds per century (mas cy$^{-1}$), phenomenologically determined by contrasting  the dynamical and measurement models of the recent ephemerides INPOP10a\cite{2011CeMDA.111..363F} to  data records spanning almost one century. For other details on such corrections, see the caption of Table \ref{ta1}.}
{\begin{tabular}{@{}ccccccc@{}} \toprule
Ephemeris & Mercury & Venus & Earth & Mars & Jupiter & Saturn \\
 & (mas cy$^{-1}$) & (mas cy$^{-1}$) & (mas cy$^{-1}$) & (mas cy$^{-1}$) & (mas cy$^{-1}$) & (mas cy$^{-1}$) \\ \colrule
INPOP10a\cite{2011CeMDA.111..363F}\hphantom{0} & $\hphantom{0}1.4\pm 1.8$ & $0.2\pm 1.5$ & $0.0\pm 0.9$ & $-0.05\pm 0.13$ & $-40\pm 42$ & $-0.1\pm 0.4$ \\
\botrule
\end{tabular} \label{ta3}}
\end{table}
\subsection{The Perihelion  of Mercury and the Sun's Angular Momentum}\lb{leti}
Although, at first sight, it might not seem immediately obvious, the correction\cite{2011CeMDA.111..363F} \eqi\Delta\dot\varpi_{\mercury}=0.4\pm 0.6~\textrm{mas~cy}^{-1}\lb{perimer}\eqf to the standard perihelion precession of Mercury determined with the INPOP10a ephemerides and reported in Table \ref{ta1} might, perhaps, hide something anomalous which deserves further investigations\cite{2012SoPh..281..815I}.

As pointed out in Section~\ref{perieli}, the relativistic dynamical models at the basis of all of the modern ephemerides are not complete in that they do not include the 1PN gravitomagnetic field of the Sun, not to say of the other major bodies of the Solar system, which causes the Lense-Thirring effect\cite{LT18}. In a coordinate system whose reference $\grf{x,y}$ plane is aligned with the equatorial plane of the central spinning body, it consists of secular precessions\cite{LT18} of the node $\Om$ and the pericenter $\omega$ of a test particle which are proportional to the angular momentum $S$ of the primary. Their magnitude is generally quite smaller than the Einstein perihelion precession; suffice it to say that they are at about the mas cy$^{-1}$ level for Mercury. In principle, such an unmodeled dynamical feature of motion  should be present in the corrections $\Delta\dot\varpi$ listed in Table \ref{ta1}. Methods mainly based on helioseismology\cite{1998MNRAS.297L..76P, 2002RvMP...74.1073C, 2013SoPh..287....9G}, which are independent of planetary dynamics, yield for the Sun's angular momentum an average value\cite{2012SoPh..281..815I}
\eqi  S_{\odot} = 1.92\times 10^{41}~\textrm{kg~m}^2~\textrm{s}^{-1}\lb{angmom}\eqf
 accurate to $\approx 1\%$. The resulting Lense-Thirring perihelion precession of Mercury is as little as\cite{2012SoPh..281..815I}
\eqi\dot\varpi_{\mercury}^{(\textrm{LT})} = -2.0~\textrm{mas~cy}^{-1}.\lb{gyro}\eqf If, on the one hand, the retrograde precession of \rfr{gyro} is about 20000 times smaller than the Einstein (prograde) precession, on the other hand, it is of the same order of magnitude of or even larger than the current level of accuracy in determining $\Delta\dot\varpi_{\mercury}$ from observations, as per Table \ref{ta1}.

Now, while \rfr{gyro} is fully compatible with the value\cite{2013AstL...39..141P, 2013MNRAS.432.3431P}  \eqi\Delta\dot\varpi_{\mercury} = -2.0\pm 3.0~\textrm{mas~cy}^{-1}\eqf determined with the EPM2011 ephemerides, it disagrees with \rfr{perimer}, obtained with the INPOP10a ephemerides, at a $4\sigma$ level. By assuming that $\Delta\dot\varpi_{\mercury}$ is entirely due to the Lense-Thirring effect, \rfr{perimer} would imply\cite{2012SoPh..281..815I} a value for the Solar angular momentum in disagreement with the several helioseismology-based determinations for it. Other possible explanations, discussed in Ref.~\refcite{2012SoPh..281..815I}, might, in principle, reside in some mutual cancelation of competing mismodeled effects impacting $\Delta\dot\varpi_{\mercury}$, or in a partial removal of the gravitomagnetic signature occurred in the estimation of some solve-for parameters. It should also be considered that, after all, \rfr{perimer} was obtained by processing just three data points of the MErcury Surface, Space ENvironment, GEochemistry, and Ranging (MESSENGER) spacecraft relative to its flybys of Mercury (two in 2008 and one in 2009); data reduction artifacts may well be possible. By the way, the teams responsible of the DE and INPOP ephemerides are currently processing  the range and Doppler data collected during the orbital phase of MESSENGER, began in early 2011, to improve, among other things, also the orbit of Mercury itself\cite{IPN014, 2014A&A...561A.115V}. More specifically, in the case of the DE430 and DE431 ephemerides\cite{IPN014}, the Hermean orbit is mainly determined by range measurements to MESSENGER. Its resulting range measurement residuals\cite{IPN014}, covering the first six months in orbit, show some signature at the Mercury orbital period that could not be removed by the current dynamical models of DE430 and DE431. The signature is due to limitations in the estimated spacecraft orbits relative to Mercury. Fig. 5 in Ref.~\refcite{IPN014} only allows to visually infer that the amplitude of such a pattern is far smaller than 50 m. Some more details are present in Ref.~\refcite{2014A&A...561A.115V} in which  1.5 years of range data to MESSENGER were analyzed to construct the INPOP13a planetary ephemerides yielding an accuracy of about\cite{2014A&A...561A.115V}  $-0.4\pm 8.4$ m in the Earth-Mercury geometric distance. Such a figure would be in agreement with the expected amplitude of the Lense-Thirring Earth-Mercury range shift\cite{2011Ap&SS.331..351I} over the same time interval calculated with \rfr{angmom}. At the time of writing, new corrections $\Delta\dot\varpi$ to the perihelion precession have not yet been determined. Whatever it is, as long as data analyses are still in progress, we feel it would be premature to draw conclusions. The MESSENGER mission is planned to be ended with a high-speed Mercury surface impact on or about 28 March 2015; see \url{http://messenger.jhuapl.edu/the_mission/extended.html} on the Internet.
\section{A Gravitational Anomaly in the Uranian System of Natural Satellites?}\lb{MAB}
As a potential new gravitational anomaly in the Solar system, it may be worthwhile reporting the case of the allegedly unexplained orbital behaviour exhibited by the inner natural satellite of Uranus Mab\footnote{From\cite{MabName} Queen Mab, a character of the Shakespearean play \textit{Romeo and Juliet}.}, also known as Uranus XXVI (26), discovered in\cite{2003IAUC.8209....1S} 2003 using the Hubble Space Telescope (HST). After its discovery, it turned out that Mab is embedded in the $\mu$-ring\cite{2006Sci...311..973S}, which was detected in 2006.

The dynamical models  successfully fit to the observations of all of the other moons of Uranus, which include also the centrifugal oblateness of the seventh planet of the Solar system and mean-motion resonances, did not work well for Mab\cite{2006Sci...311..973S}, whose residuals turned out to be unacceptably larger, being of the order of about\cite{2006Sci...311..973S,Mab011} 280 km. Since Mab is relatively bright in the data and well detached from the other moons, the measurement errors should be small. Thus, an essential part of the dynamics that determines the orbit of Mab would have been allegedly overlooked so far\cite{Mab011,2012DPS....4451308K}.  The action of a hypothetical ring of undetected moonlets in its neighborhood was proposed\cite{Mab011,2012DPS....4451308K} as a possible solution in terms of conventional gravitational physics. Further studies will be required to assess the viability of such a proposed explanation by investigating, in particular, the long-term stability of such a possible perturber ring\cite{2012DPS....4451308K}.
From the observational side, the forthcoming James Webb Space Telescope (JWST) will allow to collect valuable observations\cite{2014arXiv1403.6849T} which could help in shedding light on the alleged Mab's anomaly.

Are we facing a revival of the dichotomy which affected the perihelion of Mercury at the dawn of the Einstein's theory of gravitation? How not to note that the moonlets ring suggested by the authors of Refs.~\citen{Mab011,2012DPS....4451308K} brings to the mind the fact that one of the proposals put forth by Le Verrier\cite{LeVer1859} in 1859 to explain the Hermean anomaly  consisted just of a ring of intra-Mercurial small asteroids, named `Vulcanoids', whose searches still continue today unfruitfully\cite{2013Icar..223...48S}? Caution is in order. Indeed, should the alleged Mab orbital anomaly survive further severe theoretical and observational scrutiny,  whatever unconventional physical mechanism of gravitational origin can be devised, it will necessarily have to be put to the test on all of the other satellites of Uranus at the very least, and hopefully in other independent scenarios as well. For the time being, we limit ourselves to note that latest investigations\cite{2013MNRAS.436.3668E,2014AJ....148...76J} on the Uranian system did not deal with Mab, being mainly focussed on other satellites.

\section{The Anomalous Secular Increase of the Eccentricity of the Moon's Orbit}\lb{luna}
Modeling the motion of the Moon as accurately as required by the observations from time to time available has always represented a major challenge for physicists and astronomers since Newton's time\cite{1975intm.book.....C}; for modern overviews, see, e.g.,
Refs.~\citen{1988momo.book.....C,1998RvMP...70..589G,2010AcPSl..60..393X}. In view of the steady progresses in the Lunar Laser Ranging (LLR) technique\cite{1994Sci...265..482D}, which, in the last decades, has been able to determine the Selenian\footnote{From \grk{Sel'hnh} (`Selene'), the goddess of the Moon.} orbit at a cm level of accuracy or better\cite{2013RPPh...76g6901M} allowing for accurate tests of GTR\cite{2010LRR....13....7M}, a major limiting factor in our knowledge of the celestial course of the Moon is currently represented by a correct description of the complex geophysical processes taking place in the interior of both our planet\cite{2009iber.book.....S} and its natural satellite.\cite{2012P&SS...74...15J}

It seems that, nowadays, the Selenian orbital motion has again become a source of puzzling for scientists.
In 2009, J.~G. Williams and D.~H. Boggs, among other things, reported\cite{LLR09} also on an anomalous secular increase of the eccentricity of the Lunar orbit as large as
\eqi\Delta\dot e_{\leftmoon} = \ton{9\pm 3}\times 10^{-12}~\textrm{yr}^{-1}\lb{primorate}\eqf
determined by analyzing a multidecadal record (1970-2008) of LLR ranges with the DE421 ephemerides\cite{DE421,DE421b}. Such an effect, which could not be explained by the standard models\cite{LLR09}  of dissipative processes of tidal origin in the interiors of both the Earth and Moon, was included in Ref.~\refcite{Ande010} as one of the astrometric anomalies in the Solar system. Previous evidence for it dates back to 2003, when J.~G. Williams and J.~O. Dickey, relying upon Ref.~\refcite{2001JGR...10627933W}, quoted\cite{LLR03}
\eqi\Delta\dot e_{\leftmoon} = \ton{1.6\pm 0.5}\times 10^{-11}~\textrm{yr}^{-1}.\lb{zerorate}\eqf
A more recent analysis of an extended LLR data record (1970-2013) with the updated tidal models of the DE430 ephemerides\cite{IPN014} yielded\cite{2014PlSci...3....2W}
\eqi\Delta\dot e_{\leftmoon} = \ton{5\pm 2}\times 10^{-12}~\textrm{yr}^{-1}\lb{secondorate}:\eqf although reduced in magnitude, the anomaly is still lingering.

The authors of Ref.~\refcite{2014PlSci...3....2W} seem convinced that, sooner or later, a better modeling of the
geophysical processes of tidal origin occurring in the interiors of the Earth and Moon will
be capable to satisfactorily explain the Selenian orbital anomaly; as such, they will continue to look for conventional
physical mechanisms for it.
Nonetheless, as a complementary approach, the search for causes not related to the geophysics of the Earth and
Moon themselves is worthy of being pursued. Even if unsuccessful, it could indirectly enforce the
significance of the efforts to find an explanation in terms of standard physics. In this spirit, the author of Refs.~\citen{2011MNRAS.415.1266I,2011AJ....142...68I,2014Galax...2..259I} looked for alternative causes of $\Delta\dot e_{\leftmoon}$ in terms of Newtonian and relativistic physics, and of modified models of gravity as well. In Ref.~\citen{2011MNRAS.415.1266I}, some potentially viable gravitational effects are unsuccessfully reviewed. It turned out that several extra-accelerations arising from long-range modified models of gravity do not even induce a secular rate of change of the eccentricity. The general relativistic Lense-Thirring acceleration induced by the Earth's gravitomagnetic field on the Moon has the correct of order of magnitude, but it does not secularly affect $e$. A still undetected distant planet in the Solar system does, in principle, make $e$ cumulatively change over time, but the required mass and distance for it are completely unrealistic: suffice it to say that an Earth-sized body should be at about 30 au, while a Jovian mass should be at 200 au. In Ref.~\refcite{2011AJ....142...68I}, it was shown that an empirical acceleration proportional to the the orbiter's radial velocity $v_r$ through a multiplicative coefficient of the same order of magnitude of the current value of the Hubble parameter $H_0$ is able to cause an average rate of change of the eccentricity of the right magnitude to explain the Selenian anomaly. However, it must be pointed out that it has not yet been possible to derive such an \textit{ad-hoc} acceleration from first principles pointing towards some viable physical mechanism; in Ref.~\refcite{2013MNRAS.429..915I}, it was shown that it cannot be derived within a general relativistic cosmological framework. In Ref.~\refcite{2014Galax...2..259I}, some recently obtained effects of cosmological origin\cite{2012PhRvD..86f4004K,2013MNRAS.429..915I,2014Galax...2...13I} were unsuccessfully applied to $\Delta\dot e_{\leftmoon}$.
\section{The Faint Young Sun Paradox}\lb{fysp}
According to established evolutionary models of the Sun's history, the energy output of our star during the Archean,
from 3.8 to 2.5 Gyr ago, would have been insufficient to maintain liquid water on the Earth's surface. Instead,
there are  strong independent evidences that, actually, our planet was mainly covered by
liquid water oceans, hosting also forms of life, during that remote era.
This is the so-called\cite{1972Sci...177...52S} `Faint Young Sun Paradox' (FYSP). For a recent review of it, see Ref.~\refcite{2012RvGeo..50.2006F}
and the references therein.

More specifically,
according Ref~\refcite{1981SoPh...74...21G}, at the beginning of the Archean, i.e. 3.8 Gyr ago which corresponds to
$t_{\rm Ar} = 0.77$ Gyr in a scale based on the Zero-Age Main Sequence (ZAMS) epoch as origin of the time, the Solar luminosity\footnote{The bolometric Solar luminosity $L^{\odot}$ is a quantitative measure of the electromagnetic radiant power emitted by the Sun
integrated over all the wavelengths.} was just
\eqi L^{\odot}_{\rm Ar} = 0.75~L^{\odot}_0.\lb{LO}\eqf
Thus, by assuming the same heliocentric distance of the Earth as today, \rfr{LO} implies
\eqi I^{\odot}_{\rm Ar} = 0.75~I^{\odot}_0\lb{IrrO}\eqf for the Solar irradiance\footnote{Measured at the Earth's atmosphere, it is defined
as the ratio of the Solar luminosity to the area of a sphere centered on the Sun with a radius equal to the
Earth-Sun distance, $r$; in the following, we will assume a circular orbit for our planet.} whose current value is\cite{2011GeoRL..38.1706K}
\eqi I^{\odot}_0= 1360.8\pm 0.5~\textrm{W~m}^{-2}.\eqf
Actually, as extensively reviewed in Ref.~\refcite{2012RvGeo..50.2006F}, there is a wide and compelling body of evidence that the Earth hosted liquid
water, and even life, during the entire Archean spanning about 1.3 Gyr. As such, our planet could not be
entirely frozen during such an era, as, instead, it would have necessarily been if it really received
only about $75\%$ of the current Solar irradiance, as provided by \rfr{IrrO}.
So far, studies aiming to resolve the puzzle of the FYSP have been chiefly the prerogative of geophysicists, heliophysicists and paleoclimatologists\cite{2004LRSP....1....2W,2007LRSP....4....3G,2008AsBio...8.1127H,2010Natur.464..744R,Driese20111,2012ApJ...760...85C,2013EOSTr..94R..76S,2013Sci...339...44K,2013Sci...339...64W,2014P&SS...98...77K}. Despite all the efforts lavished so far, the FYSP is still  away from being satisfactorily resolved\cite{2010Natur.464..687K,2011Natur.474E...1G}: in fact, it even seems that it is escalating\cite{2012GeoRL..3923710K}.

An alternative-and minority so far-view\cite{2000GeoRL..27..501G,2007ApJ...660.1700M,2012RvGeo..50.2006F,2013Galax...1..192I,2014arXiv1405.4369S} of the FYSP implies a steady recession of the Earth's orbit during the entire Archean eon from a closer location to its present-day heliocentric distance in such a way that the Sun's luminosity, whose evolutionary history is rather well established, allowed for the existence of the liquid water throughout such a putative migration. Some models set the limiting value of the Solar irradiance to keep liquid oceans on the Earth's surface at about\cite{1997GPC....14...97L} \eqi I^{\odot}_{\rm oc}\approx 0.82~I^{\odot}_0;\eqf if $I^{\odot}$ was really equal to $I^{\odot}_{\rm oc}$ at the beginning of the Archean, then, at that time, the Earth should have been necessarily closer to the Sun than now by about\cite{2013Galax...1..192I} $4.4\%$. If so, some physical mechanism should have subsequently displaced the Earth to its current location in the next 1.3 Gyr, when, at the beginning of Proterozoic ($t_{\rm Pr}=2.07$ Gyr in the ZAMS time scale), $L^{\odot}$ would have reached the limiting value $L^{\odot}=0.82~L^{\odot}_0$ according to the Sun's luminosity models. It turns out\cite{2013Galax...1..192I} that the constraint \eqi I^{\odot}(t)\approx I^{\odot}_{\rm oc}\approx 0.82 I^{\odot}_0\eqf from $t_{\rm Ar}$ to $t_{\rm Pr}$ imposes a relative recession rate
\eqi\rp{\dot r}{r}\approx \rp{1}{7 t_0\ton{1-\rp{2}{7}\rp{t}{t_0}  }} \approx 3.4\times 10^{-11}~\textrm{yr}^{-1},~t_{\rm Ar}\leq t\leq t_{\rm Pr} \lb{rrate}\eqf where $t_0=4.57$ Gyr is the present epoch in the ZAMS time scale. In Refs.~\citen{2000GeoRL..27..501G,2007ApJ...660.1700M,Minton012}, some more or less potentially viable mechanisms to explain the putative recession of the terrestrial orbit in terms of standard gravitational physics and Sun's mass loss at the birth of its life  were unsuccessfully proposed.

In Ref.~\refcite{2013Galax...1..192I}, after having ruled out a cosmological mechanism based on the currently accepted $\Lambda$CDM (Lambda Cold Dark Matter) scenario, the FYSP was viewed, for the first time, as a possible gravitational anomaly since the putative shift of the heliocentric distance of the Earth was attributed to a certain long-range modified model of gravity\cite{2013PhRvD..87d4045P,2014CQGra..31h5003I}. Indeed, a certain class\cite{2013PhRvD..87d4045P} of modified gravitational theories with nonminimal coupling between the matter and the gravitational field, which, among other things, predict also violations of the equivalence principle, yields an extra-acceleration which induces a long-term variation of the distance between the components of a localized, gravitationally bound two-body system\cite{2014CQGra..31h5003I}. If, on the one hand, it is not an \textit{ad-hoc} explanation since
it is rooted in a well-defined theoretical framework, on the other hand, further work is still required to fully elucidate several key aspects of this model. Moreover, it should somehow be put to the test independently in different systems. Later, the authors of Ref~\refcite{2014arXiv1405.4369S} suggested that a change in the Newtonian gravitational constant, in alleged agreement with  present-day bounds inferred from several independent sources, could accommodate the FYSP. The percent change $\Delta G/G$ postulated in Ref~\refcite{2014arXiv1405.4369S} can be translated, with some caution, to a secular decrease as large as\cite{2014arXiv1405.4369S}
\eqi\rp{\dot G}{G}\simeq -4\times 10^{-12}~\textrm{yr}^{-1}.\lb{rateG}\eqf Actually, the rate of \rfr{rateG} is too large by about two orders of magnitude with respect to the most recent bounds determined from Solar system's planetary
dynamics\cite{2013MNRAS.432.3431P,2013AstL...39..141P,2014arXiv1409.4932F}, and by about one order of magnitude if latest constraints from LLR are considered\cite{2004PhRvL..93z1101W,2007CQGra..24r4533M}.
\section{The Secular Decrease of the Mass Parameter \textit{GM} of the Sun}\lb{massa}
Recent works\cite{2012SoSyR..46...78P,2013MNRAS.432.3431P,2014CeMDA.119..237P} in the field of Solar system's planetary dynamics by the EPM  team have lead to some reciprocally compatible estimates of a statistically non-zero variation of the Solar mass parameter defined as \eqi\mu_{\odot} \doteq G M_{\odot};\eqf  they are listed in Table \ref{ta4}.
\begin{table}
\tbl{Relative rates of change $\dot\mu_{\odot}/\mu_{\odot}$, in yr$^{-1}$, of the Solar mass parameter $\mu_{\odot}\doteq GM_{\odot}$ determined with the EPM2008\cite{Pit09}, EPM2010\cite{2012SoSyR..46...78P} and EPM2011\cite{2013MNRAS.432.3431P} ephemerides.}
{\begin{tabular}{@{}ccc@{}} \toprule
EPM2008\cite{2014CeMDA.119..237P} & EPM2010\cite{2012SoSyR..46...78P} & EPM2011\cite{2013MNRAS.432.3431P}  \\
(yr$^{-1}$) & (yr$^{-1}$) & (yr$^{-1}$)  \\ \colrule
$\ton{-5.9\pm 4.4}\times 10^{-14}$ & $\ton{-5.0\pm 4.1}\times 10^{-14}$ & $\ton{-6.3\pm 4.3}\times 10^{-14}$ \\
\botrule
\end{tabular} \label{ta4}}
\end{table}
Despite their presently low level of statistical significance, such determinations are, in principle, important since they have been obtained from about one century  of various kinds of planetary positional observations; as such, the estimates in Table \ref{ta4} are independent of any heliophysical model.

In principle, a change in $\mu_{\odot}$ can be due to either the Newtonian gravitational constant $G$ and/or the mass $M_{\odot}$ of our star; since independent analyses performed with the LLR technique\cite{2004PhRvL..93z1101W,2007CQGra..24r4533M} did  not yield net changes in $G$, the results in Table \ref{ta2} should be attributed to a Sun's mass loss. Known physical mechanisms for it are electromagnetic radiation generated in the core nuclear burning ($80\%$) and average Solar wind ($20\%$). A recent figure for the Sun's total mass loss, obtained from heliophysics measurements during the 11-year Solar cycle and cited in Ref.~\refcite{2014arXiv1409.4932F}, is\cite{2011ApJ...737...72P}
\eqi\rp{\dot M_{\odot}}{M_{\odot}}= \ton{-5.5\pm 1.5}\times 10^{-14}~\textrm{yr}^{-1}.\lb{varM}\eqf The range of values in \rfr{varM} is compatible with those reported in Table \ref{ta4}, but it is able to explain only about $35-36\%$ of them. The authors of Ref.~\refcite{2013MNRAS.432.3431P} warns that  a reliable assessment of asteroidal and cometary matter falling on the Sun should be taken into account in correctly evaluating the overall rate of mass decrease of the Sun, but they did not provide any estimate for such a contribution.
More recent analyses with the INPOP13c\cite{2014arXiv1409.4932F} ephemerides did not confirm the findings reported in Table \ref{ta4} since the values obtained for $\dot\mu_{\odot}/\mu_{\odot}$ are statistically compatible with zero.

Further heliophysics-independent, dynamically inferred constraints on the decrement of the Solar mass are worth of being further studied since, at least in principle, they might open up new windows on the role of Dark Matter in the physical processes taking place in the interior of our star. Such an appealing possibility might have gained some credit after the recently published preliminary results concerning an alleged detection of axions  supposedly emitted by the Sun\cite{2014xru..confE..74F}.
\section{The Flyby Anomaly}\lb{flyby}
As `Flyby Anomaly' (FA),\cite{2007NewA...12..383A,2008PhRvL.100i1102A,2009PhT....62j..76N,2009SSRv..148..169T,Ande010,2013AGUFMSM33B2187A}
it is intended the collection of unexplained increases $\Delta v_{\infty}$ in the asymptotic line-of-sight velocity $v_{\infty}$, of the order of $\approx 1-10$ mm s$^{-1}$ with uncertainties as little as $\approx 0.05-0.1$ mm s$^{-1}$,  experienced by the interplanetary spacecraft Galileo\footnote{At its first Earth gravity assist on 8 December 1990; two, years later, at its second Earth flyby, the anomaly was not detected.}, NEAR, Cassini\footnote{For it, the anomaly amounted to\cite{2008PhRvL.100i1102A} $2\pm 1$ mm s$^{-1}$.}, Rosetta\footnote{At its first Earth gravity assist on 4 March 2005; the anomaly did not show up at its second (13 November 2007) and third (13 November 2009) Earth flybys.\cite{2010arXiv1006.3555B,Biele012}} and, perhaps, Juno\cite{2013AGUFMSM33B2187A,SDaily} at their Earth flybys. The FA has not yet been detected when such spacecraft flew by other planets,  perhaps due to their still relatively inaccurate gravity field models\cite{2007arXiv0711.2781B,2009SSRv..148..169T,2010arXiv1006.3555B}.  In all of the reported flybys, all the efforts of either the navigation and the radio science teams to find an explanation of this anomaly were unsuccessful. Juno flew by the Earth on 9 October 2013; in principle, it would be an ideal probe to detect the anomaly, if any, since the control sequence for the spacecraft did not introduce  significant translational forces over an 8-days interval centered on the perigee passage\cite{2013AGUFMSM33B2187A}. Data analysis are still ongoing, but, according to some sources\cite{SDaily},  a discrepancy between the measured and the predicted asymptotic speeds would have occurred.

The authors of Ref.~\refcite{2008PhRvL.100i1102A} proposed the following empirical formula
\eqi \rp{\Delta v_{\infty}}{v_{\infty}}= \rp{2\Omega_{\oplus} R_{\oplus}}{c}\ton{\cos\delta _{\rm in} - \cos\delta_{\rm out}}\lb{formula},\eqf
where $\Omega_{\oplus},~R_{\omega}$ are the Earth's angular velocity and equatorial radius, respectively, and $\delta_{\rm in},~\delta_{\rm out}$ are the declinations of the incoming and outgoing osculating asymptotic velocity
vectors. So far, it has not yet been possible to derive \rfr{formula} consistently from first principles within some theoretical frameworks; to this aim, it is interesting to note that the Newtonian gravitational constant $G$ does not enter \rfr{formula}, as if it did not describe a truly gravitational effect. If, on the one hand, \rfr{formula}, which does not contain free, adjustable parameters, was able to explain the flybys occurred when Ref.~\refcite{2008PhRvL.100i1102A} was published, on the other hand it failed to account for the null result occurred at the second and third Rosetta flybys.\cite{2010arXiv1006.3555B,Biele012} In the case of Juno's flyby, \rfr{formula} predicts a shift as large as\cite{2013AGUFMSM33B2187A} 7 mm s$^{-1}$.
In Ref.~\refcite{2007arXiv0711.2781B}, an alternative empirical formula, which contains three arbitrary parameters, was successfully applied to the anomalous flybys known at that time. Essentially, it would be a consequence of a modification of the usual Earth's Newtonian gravitational monopole due to an alleged preferred-frame effect with respect to the CMB, not modeled within the standard PPN framework. Later, it was shown to be in agreement with the observed null anomalies of the other Rosetta flybys of Earth\cite{2010arXiv1006.3555B}; its predicted anomaly\cite{2013arXiv1312.1139B} for the Juno's flyby has the opposite sign with respect to that provided by \rfr{formula}.

Several more or less potentially viable explanations of the FA  in terms of either gravitational (standard\cite{2008ASSL..349...75L,2009ScReE2009.7695I,2010cosp...38.3845H,2010JGCD...33.1115A,2014AdSpR..54.2441I} and modified\cite{2007arXiv0711.2781B,2008MNRAS.389L..57M,2009PhRvD..79b3505A,2010IJMPA..25..815C,2010IJMPA..25.4577A,2013IJMPA..2850074A,2014GReGr..46.1741V,2014AdSpR..54..788A,2014PhLA..378.3007P}) and non-gravitational\cite{2011AnP...523..439R} effects have been put forth in the course of latest years, but none has proved itself fully satisfactorily so far. Some authors suggested to test it with existing or proposed space-based missions such as the GNSS constellation\cite{2012IJMPD..2150035B} and STE-QUEST\cite{2013P&SS...79...76P}. In general, a major problem which all non-standard gravitational explanations  must face is the the need of being  tested independently of the FA itself, and to explain why their consequences generally do not manifest also in the motion of Earth's artificial satellites moving along bound orbits.
\section{The Pioneer Anomaly}\lb{Pio}
At the end of the twentieth century, it was reported\cite{1998PhRvL..81.2858A,2002PhRvD..65h2004A} that radio tracking data from the Pioneer 10 and 11 spacecraft exhibited a small anomalous blue-shifted frequency drift uniformly changing with a rate  \eqi\Delta\dot\nu=\ton{5.99\pm 0.01}\times 10^{-9}~\textrm{Hz~s}^{-1}\eqf that was interpreted as a constant and uniform deceleration approximately directed towards the Sun
\eqi  A_{\rm Pio} = \ton{8.74\pm 1.33}\times 10^{-10}~\textrm{m~s}^{-2},\lb{Apio}\eqf at heliocentric distances $\approx 20-70$ au. From then onwards, such an alleged violation of the Newtonian inverse-square law of gravitation has been known by the denomination of `Pioneer Anomaly' (PA).
Later, its existence was repeatedly confirmed by further independent analyses by other individuals\cite{2002gr.qc.....8046M,2007A&A...463..393O,2009IJMPD..18..717T} and teams of researchers\cite{2009AdSpR..43.1538L}.
In the absence of any physical theory  predicting \rfr{Apio}, the primary candidate remained some systematic biases generated by spacecraft systems themselves. However, with the data available at that time, neither the authors of Refs.~\citen{1998PhRvL..81.2858A,2002PhRvD..65h2004A} nor anyone else were able to find a viable non-gravitational effect that was both large enough and constant enough to explain the PA. Subsequent years witnessed a steady flow of papers proposing a variety of conventional and exotic physical mechanisms, of both gravitational and non-gravitational nature, to accommodate the PA. In view of their exceptionally large number and inhomogenous level of soundness, it is not possible to keep track of all of them here; see, e.g., Refs.~\citen{2002PhRvD..65h2004A,2010LRR....13....4T} and references therein. In view of the lingering inability of finding satisfactory explanations to the PA in terms of both standard and unconventional physics over the years,  a comprehensive new investigation of the anomalous behavior of the two probes was launched since mid of 2006 after the recovery of the much-extended set of radio-metric Doppler data for both spacecraft in conjunction with the
newly available complete record of their telemetry files and a large archive of original project documentation\cite{2006IJMPD..15....1T,2006CaJPh..84.1063T,2008AIPC..977..264T,2009SSRv..148..149T}.

The fruits of such hard and heroic efforts were not long in coming.
Indeed, by analyzing the new data,  conviction that suitably designed thermal models could likely explain the PA\cite{2008PhRvD..78j3001B,2009NJPh...11k3032R,2009SSRv..148..149T,2009PhRvD..79d3011T,2010AcAau..66..467R,2010SSRv..151...75B,2010SSRv..151..123R,2011AnP...523..439R,2012PhLB..711..337F}
began to make its way. In 2012, it was shown\cite{2012PhRvL.108x1101T} that an appropriate model of the recoil force associated with an anisotropic emission of thermal radiation off the spacecraft is able to accommodate about $80\%$  of the unexplained acceleration plaguing the telemetry of both the Pioneer probes as far as its magnitude, temporal behavior, and direction are concerned. The remaining $20\%$ does not represent a statistically
significant anomaly in view of the uncertainties in the acceleration
estimates using Doppler telemetry and thermal models. The authors of Ref.~\refcite{2012PhRvL.108x1101T} concluded that the anomalous acceleration of the Pioneer spacecraft is consistent with known physics. The same results were obtained almost contemporaneously in the same year by an independent study\cite{Berto012} on the thermal effects of the Pioneer probes. The same conclusions were drawn also in 2014 by a further independent analysis\cite{2014PhRvD..90b2004M} of the navigation telemetry with a thermal finite element model of the spacecraft achieving orbital solutions that do not require the addition of any anomalous acceleration other than that of thermal origin.

On the other hand, the fact that the PA was unlikely due to some exotic gravitational mechanism external to the spacecrafts, at least in the form of a constant and uniform acceleration directed towards the Sun, began somehow to creep into people's knowledge  some years before, when systematic investigations about its putative effects on bodies other than the Pioneer probes themselves began to be performed since 2006. Indeed, apart from some earlier studies\cite{2002PhRvD..65h2004A} focussed only on the inner planets of the Solar system and a few preliminary analyses on the impact of a PA-type force on long-period\footnote{Such bodies are not ideal probes for testing gravitational theories because of several perturbing non-gravitational forces\cite{2012A&A...548A..81M} like out-gassing as they approach the Sun.} comets\cite{comets02} and the configuration of the Oort cloud\cite{2003Icar..165..219W}, the basic question if the PA manifests itself also on the major bodies of the Solar system orbiting in those regions traversed by the Pioneer probes seems to have been substantially neglected for a long time. By the way, in 2006, the authors of the unpublished Ref.~\refcite{2006gr.qc.....4068T} explicitly wondered why the anomalous acceleration toward the Sun is found only for spacecrafts, but not for the trajectories of planets. Such kind of investigations, complementary to those looking explicitly for some non-gravitational conventional mechanisms and indirectly enforcing them, initiated systematically in 2006 with the publication of Ref.~\cite{2006NewA...11..600I}. In it, simulated residuals of certain functions of  $\alpha$ and $\delta$ of the two farthest gaseous planets of the Solar system and the dwarf planet Pluto induced by a constant and uniform acceleration with the same magnitude of the PA and pointing towards the Sun were preliminarily compared to existing residuals of the same bodies produced with some planetary ephemerides constructed without modeling the PA. It turned out that the PA-induced anomalous signatures of Uranus, Neptune and Pluto would be far too large, even by considering the possibility that part of them could have been somewhat removed from the real residuals in the estimation of, say, the initial conditions.
Such a conclusion was essentially confirmed some years later, when independent analyses by the DE\cite{2008AIPC..977..254S,Folkner09,2010IAUS..261..179S} and INPOP\cite{2009sf2a.conf..105F,Fienga010,PitGioi,2012sf2a.conf...25F} teams were performed by explicitly modeling a PA-type Sunward extra-acceleration and fitting the resulting \textit{ad-hoc} modified dynamical models to the same data records of the standard ephemerides. Meanwhile, several papers on the same topic had appeared in the literature, dealing sometimes with either other bodies as probes and different forms of the PA itself. In Ref.~\refcite{2006ApJ...642..606P}, it was suggested to look at the Trans-Neptunian Objects (TNOs) to confirm or refute the existence of the PA as a real effect independent of the spacecraft. Such a suggestion was implemented in Ref.~\refcite{2007ApJ...666.1296W} in which observations to a selected sample of TNOs from 20 to 100 au were processed with modified dynamical models including a PA-type acceleration $A$ by finding $A=\ton{0.87\pm 1.6}\times 10^{-10}$ m s$^{-2}$, a result which is statistically compatible with zero. The authors of Ref.~\refcite{2006JSpRo..43..806R}
parameterized the PA in terms of a change of the effective reduced Sun's mass felt by Neptune by finding it nearly two orders of magnitude
beyond the current observational constraint. Moreover, they noted that the Pioneer 11 data contradict the existing Uranus ephemerides, obtained
without explicitly modelling the PA, by more than one order of magnitude. In Ref.~\refcite{2007FoPh...37..897I}, some long-range modified gravity models tuned to predict a PA-type extra-acceleration were unsuccessfully put to the test by comparing their predicted effects with existing residuals of Uranus, Neptune and Pluto. The same occurred also for the perihelion precessions induced by a putative PA-like anomalous acceleration, worked out in Ref.~\refcite{2006NewA...11..600I}, which were confronted\cite{2007FoPh...37..897I} with corrections $\Delta\dot\omega$ determined for Uranus, Neptune and Pluto in Ref.~\refcite{PitPio}. Incidentally, also the author of Ref.~\refcite{PitPio} drew the same conclusion. In Ref.~\refcite{2007PhRvD..76d2005T}, while supporting the previous negative findings about the existence of a gravitational PA extra-acceleration acting on orbiting bodies in the outer regions of the Solar system on the basis of their existing residuals, it is argued that Neptune could not rule out it because of the insufficient accuracy of its observations
In Ref.~\refcite{2009ApJ...697.1226P}, modified dynamical models including an additional PA-like acceleration were fitted to data records for
Uranus, Neptune and Pluto showing that the Plutonian ephemerides available at that time did not preclude the existence of the Pioneer effect because of their relatively poor accuracy. Some velocity-dependent  forms of the PA, discussed in Ref.~\refcite{2008AIPC..977..254S}, were unsuccessfully applied\cite{2009IJMPD..18..947I} to the orbital motions of Uranus, Neptune and Pluto by comparing their theoretically predicted perihelion precessions to corrections $\Delta\dot\varpi$ determined with the EPM2006\cite{2008IAUS..248...20P} ephemerides without explicitly modeling the PA. In Ref.~\refcite{2010MNRAS.405.2615I}, the search for a putative gravitational Pioneer effect moved to the Neptunian system of natural satellites, whose existing residuals were inspected also in this case without success. Finally, going back to the early works, it is to be noted that the authors of Ref.~\refcite{2003Icar..165..219W} found strong tensions between the established evidence that the Galactic tide is dominant in making Oort cloud comets observable and the action of a putative PA-like acceleration in those remote peripheries of the Solar system, as independently confirmed later in Ref.~\refcite{2012MNRAS.419.2226I}.
All such studies mainly dealt with the `standard' form of the PA viewed as a constant and uniform acceleration. In  Ref.~\refcite{2011PhRvL.107h1103T}, an analysis concerning a temporally varying behavior of the PA was reported. Also in this case, its alleged impact on planetary orbital motions was ruled out by a comparison with the observations\cite{2012MPLA...2750071I}.
As outlined in Section~\ref{perieli}, dedicated planetary missions\cite{2012ExA....33..753A,2012ExA....34..203C,2014P&SS..104...93T} to the outermost icy gaseous giants of the Solar system would allow to greatly improve their orbital determination, thus allowing for much more stringent tests of the PA-and, more generally, of gravitational theories-in the far regions of the Sun's realm.

Despite the growing evidence towards a satisfactorily solution of the PA in terms of conventional non-gravitational effects peculiar to the two spacecraft, there are still researchers continuing to look for new gravitational physical mechanisms\cite{2012PhRvD..86f4004K,2012PRI..2012E...1V,2012PhRvD..86f4023A,2012PhRvD..85h4017A,2013CEAS....5...19K,2013AdSpR..51.1266C,2013PLoSO...878114F,2013Ap&SS.347..235M} able to accommodate the PA. A positive feature of such attempts is that they are now explicitly dealing with the need of explaining why the PA does not show up in the orbital motions of the Solar system's bodies moving in the same regions where the anomaly manifested itself, a crucial issue that cannot be set aside.

Finally, it is worthwhile mentioning that, over the years, several interesting concept studies  about dedicated space-based missions  to test the PA were proposed\cite{2002IJMPD..11.1545A,2005gr.qc.....6139T,2005ESASP.588....3D,2006JSpRo..43..806R,2007IJMPD..16.1611B,2007AdSpR..39..291T}. Among them, it is of particular interest the one\cite{2008PhLB..659..483N,2009SSRv..148..149T} aiming to use data from New Horizons, which should reach Pluto  in July 2015.
\section{The Anomalous Secular Increase of the Astronomical Unit}\lb{UA}
At the beginning of the present century, an anomalous secular increase of the astronomical unit of the order of
$\dot{\textrm{au}}=0.15\pm 0.04~\textrm{m~yr}^{-1}$ was reported\cite{2004CeMDA..90..267K}. As shown by subsequent independent investigations\cite{2005tvnv.conf..163S,Ande010,Pit012}, its status  has remained somewhat controversial so far. As in the case of the PA and, to a lesser extent so far, the FA, also such an alleged anomaly motivated many researches to find viable explanations in terms of various either conventional or unconventional physical mechanisms\cite{2005JCAP...09..006I,2006PhRvD..74l4006M,2007CQGra..24.5031M,2008EPJST.163..255L,2008ApJ...679..675V,2008ASSL..349...75L,2009SSRv..148..501L,2009PASJ...61.1373I,2009NewA...14..264A,2010AdSpR..45.1007A,2009PASJ...61.1247M,2011GReGr..43.2127A,2011AJ....142...68I,2011ChPhC..35..914L,2012JApA...33..201A,2013AdSpR..52.1297A,2013Ap&SS.347...41W}.

The various attempts\cite{2004CeMDA..90..267K,2005tvnv.conf..163S,Ande010,Pit012} to investigate it have been performed so far within the framework of the International Astronomical Union (IAU) 1976 System of Astronomical Constants, available on the Internet at \url{http://www.iau.org/static/resolutions/IAU1976_French.pdf}, which, in turn, is reminiscent of pre-relativistic eras when, among other things, only relative distances (expressed in astronomical units) in the Solar system could be determined since absolute ones could not be estimated with high accuracy. More precisely, the IAU 1976 system prescribed to measure the masses in units of Solar masses, the durations in days $D$ (made of 86400 seconds), and the distances in  astronomical units of length. The definition of the astronomical unit of length (i.e. the astronomical unit) is based on the value of the Gaussian gravitational constant $k$ through the non-relativistic third Kepler's law\cite{Pit012}. Indeed, according to the IAU 1976 definition, the astronomical unit of length is that length (A) for which the Gaussian gravitational constant ($k$), which was a defining constant, takes the value of $0.01720209895$ when  the astronomical units of length, mass and time are used. In view of the third Kepler's law, the dimensions of $k^2$ are those of the Newtonian gravitational constant, i.e. $[k^2]=\textrm{L}^3~\textrm{M}^{-1}~\textrm{T}^{-2}$. Now, the value of the astronomical unit in the Syst\`{e}me International (SI), i.e. in metres, was determined observationally from fits to planetary ephemerides with a certain uncertainty in such a way that the value in SI units of the Heliocentric gravitational constant (or Sun mass parameter) $\mu_{\odot}$ had to be derived from $k$, along with the adopted value for the astronomical unit in metres through the third Kepler's law in the form \eqi \mu_{\odot}= \textrm{A}^3 k^2 D^{-2}.\eqf Thus, the SI value of the astronomical unit of mass turned out to be dependent on the SI value of the astronomical unit of length.

Recent improvements in Solar system ephemerides, fully based on relativistic models for both orbital motions and propagation of electromagnetic waves, made such a scenario obsolete and misleading\cite{2008A&A...478..951K,Cap09,Cap011} also because modern observations like planetary radar ranging, spacecraft observations, Very Long Baseline Interferometry (VLBI), etc. allow nowadays for very accurate absolute measurements of distances, thus making relative measurements unnecessary. Moreover, in current ephemerides, the Sun gravitational parameter is directly determined in SI units with such an accuracy that even tiny temporal variations of it have become accessible (See Section~\ref{massa}). It is clear that, with the IAU 1976 definition of the astronomical unit, the temporal variability of $\mu_{\odot}$ would make the astronomical units of mass and length time-dependent quantities which are unsuitable for their original purposes.

For all such reasons, as proposed, e.g., in Ref.\refcite{Cap012}, in the Resolution B2 of the XXVIII IAU General Assembly, available on the Internet at \url{http://syrte.obspm.fr/IAU_resolutions/Res_IAU2012_B2.pdf},  it was decided to redefine the astronomical unit to be a conventional unit of length expressed in a defining number of SI meters, so that now it is
$1~\textrm{au} = 149,597,870,700~\textrm{m}$  exactly. As a consequence, the Gaussian constant $k$  does not have a role any more: it was deleted from the IAU System of astronomical constants. Moreover, the experimental determination of the astronomical unit in SI units was abandoned, and the SI value of $\mu_{\odot}$ is currently determined experimentally.
\section{Summary and Conclusions}\lb{fine}
Anomalies in the standard behaviour of natural and artificial systems in the Sun's realm as expected on the basis of conventional physics and known matter-energy distributions have, in principle, a great potential to uncover modifications in our currently accepted picture of natural laws. Nonetheless, before this dream really comes true, it is mandatory that the unexpected patterns are confirmed to an adequate level of statistical significance by independent analyses, and any possible conventional viable mechanism which, to the best of our knowledge, could be responsible of them is reliably excluded. They are not  easy tasks to be accomplished, requiring often time, resources and relentless efforts. Moreover, even when a reasonable agreement is reached about the true existence of an anomaly, the search for gravitational mechanisms, either rooted in known effects or in alternative theories, cannot be separated from devising independent tests of them, which must be applied also to systems other than those for which they were originally proposed. If such minimal requirements are met,it is always worth following new tracks since even negative results obtained within a given research program can serendipitously and unexpectedly turn out to be valid in different ones, leading also to technological advancements. Below, we offer a concise summary of the more or less established anomalies examined in this review.
\begin{itemize}
\item \textbf{Anomalous perihelion precessions (Section~\ref{perieli})}. After the appearance of the alleged anomalous perihelion precession of Saturn, which lead a discussion for a while, most recent data analyses yielded corrections to the Kronian periehlion rate statistically compatible with zero. At present, according to the EPM team, there are non-zero precessions for Venus and Jupiter, but their best estimates are only $1.6-2$ times larger than the quoted uncertainties. Based on the three Mercury flybys by MESSENGER, in 2011, the INPOP team determined a range of allowed values for any Hermean anomalous perihelion precession which, if on the one hand, are compatible with zero, on the other hand imply a value of the Sun's angular momentum significantly smaller than its currently accepted values mainly inferred with helioseismology. The completion of the MESSENGER mission in 2015 and the analysis of all its radiometric data will be crucial in notably improving the orbit of the first planet of the Solar system and to test the Lense-Thirring effect. A further step will occur with the planned BepiColombo mission, to be launched in 2016. Our knowledge of the orbit of Pluto, currently known only from optical data of relatively modest accuracy, will greatly benefit from accurate tracking of the New Horizons spacecraft, which will reach the Plutonian system in 2015. As far as Uranus and Neptune are concerned, it will be necessary to wait and see if some of the recently proposed missions like, e.g., Uranus Pathfinder and OSS, will be finally approved. It is highly  desirable that the teams of astronomers currently engaged in the production of Solar system ephemerides of ever increasing accuracy determine corrections to the standard rates of change of all the orbital elements of all the planets in the same way as they have done so far for the perihelia. This would allow to put to the test several classes of gravitational theories predicting non-spherically symmetric corrections to the inverse-square law.
 \item \textbf{Mab orbital anomaly (Section~\ref{MAB})}. After the discovery in 2003 of such a tiny natural moon of Uranus, it was pointed out that the dynamical models used to successfully determine the orbits of the other Uranian satellites did fail with it, leaving huge unexplained residuals as large as about 280 km. A conventional, Vulcanoids-type explanation has been proposed so far in terms of a not yet discovered ring of moonlets whose long-term stability is currently theoretically investigated.
    In the event that one were to consider the orbital behaviour of Mab as a truly gravitational anomaly worthy of investigations, it must be bear in mind that any gravitational mechanism that can be devised must necessarily be put to the test independently in other systems as well.  A future observational survey with the JWST telescope will be of great help.
    \item \textbf{The anomalous secular increase of the Lunar eccentricity (Section~\ref{luna})}. Despite increasingly accurate models of the geophysical processes of tidal origin taking place in the interiors of the Earth and Moon, this small unexplained secular rate of the order of $\approx 10^{-12}$ yr$^{-1}$ is still lingering. Space geodesists and geophysicists are convinced that, sooner or later, it will disappear as an anomaly thanks to better modeling, but, until that happens, it is quite admissible to look for alternative explanations which, if formulated as modifications of the standard laws of gravity, have necessarily to cope with, e.g., the motions of the plethora of the Earth artificial satellites as well.
    \item \textbf{The Faint Young Sun Paradox (Section~\ref{fysp})}. So far, such a puzzle, which seems to be still far from being satisfactorily resolved in terms of conventional physics, has not been considered as a gravitational anomaly worthy of an explanation in terms of new physics. Indeed, it has always been the playground of heliophysicists, atmospheric physicists, climatologists, archeoastronomers. Nonetheless, recent researches showed that it might be considered as a test bench also for alternative models of gravity predicting new orbital effects. In principle, it could  be the topic of Dark Matter researches as well, in view of putative interactions of such an elusive and exotic form of matter within the interior of the Sun itself.
    \item \textbf{The secular decrease of the gravitational constant $GM_{\odot}$ of the Sun (Section~\ref{massa})}. The EPM team recently claimed to have measured it, although the associated uncertainty is still large. The INPOP team did not confirm it. A model-independent, dynamical determination of the temporal behaviour of such a key Solar parameter is certainly important to constrain heliophysical models and, in principle, also the role that Dark Matter may play in them, if any.
    \item \textbf{The Flyby Anomaly (Section~\ref{flyby})}.
     It has been detected so far only in the gravitational field of the Earth in occasion of certain flybys by interplanetary spacecraft (Galileo, NEAR, Cassini, Rosetta, and, perhaps, Juno) in search of the required gravity assists to reach their final targets. In some other Earth flybys, it did not manifest itself. No doubts exist on the existence of such an anomaly since its level of statistical significance is high in almost all its detections. Until now, it has always resisted any attempt of explanation. It does not seem that the models of non-gravitational thermal forces successfully used for the Pioneer Anomaly work well also for this anomaly.
    \item \textbf{The Pioneer Anomaly (Section~\ref{Pio})}.
    From 2012 onwards, the best known and perhaps longest-lasting of all of the alleged Solar system anomalies seems to have finally found satisfactory explanations in terms of conventional physical mechanisms pertaining the thermal effects which affected the Pioneer probes. On the other hand, a growing body of  more or less compelling evidence against its putative gravitational nature  had been accumulating in previous years when it was shown that it does not affect the orbits of the major (and, perhaps, also minor) bodies of the Solar system moving in the same regions in which the anomaly manifested itself. Nonetheless, there are still researchers looking for unconventional solutions in terms of fundamental physics.
    \item \textbf{The anomalous secular increase of the astronomical unit (Section~\ref{UA})}.  The claimed detections of a steady increment of the unit of length in the System of Astronomical  Constants, which during some recent years had earned a certain resonance, were all obtained on the basis of its old definition based on the pre-relativistic third Kepler law through the Gaussian gravitational constant. After 2012, the latter one was cancelled from the list of astronomical constants, and the astronomical unit was promoted to the rank of defining constant.
\end{itemize}

\bibliographystyle{ws-ijmpd}
\bibliography{invrevijmpd_bib}
\end{document}